\documentclass[11pt,letterpaper]{article}

\usepackage[latin1]{inputenc}
\usepackage[T1]{fontenc}
\usepackage{amsfonts}
\usepackage{amssymb}
\usepackage{color}
\usepackage{fullpage}
\usepackage[dvips]{graphicx}
\usepackage{amsmath,amsthm}
\usepackage{latexsym}

\newtheorem{prop}{Prop}
\newtheorem{claim}{Claim}
\newtheorem{fact}{Fact}
\newtheorem{defi}{Definition}

\newtheorem{lemma}{Lemma}
\newtheorem{corol}{Corollary}

\newtheorem{theo}{Theorem}

\newcommand{\cadre}[1]
{
\begin{tabular}{|p{13cm}|}
\hline
#1 \\
\hline
\end{tabular}
}

\newcommand{\COMMENT}[1]{}
\newcommand{\ket}[1]{|#1\rangle}

\def\01{\{0,1\}}

\begin{document}

\title{The role of help in Classical and Quantum Zero-Knowledge}
\author{Andr\'e Chailloux$^*$ \\
LRI\\
Universit\'e Paris-Sud\\
andre.chailloux@ens-lyon.org\\
\and
Iordanis Kerenidis\thanks{Supported in part by ACI Securit\'e Informatique SI/03 511 and ANR AlgoQP grants of the French Ministry and in part by the European Commission under the Intergrated Project Qubit Applications (QAP) funded by the IST directorate as Contract Number 015848.}\\
CNRS - LRI\\
Universit\'e Paris-Sud\\
jkeren@lri.fr
}

\maketitle \thispagestyle{empty}

\begin{abstract}
We study the role of help in Non-Interactive Zero-Knowledge protocols and its relation to the standard interactive model. In the classical case, we show that help and interaction are equivalent, answering an open question of Ben-Or and Gutfreund (\cite{BG03}). This implies a new complete problem for the class SZK, the Image Intersection Density. For this problem, we also prove a polarization lemma which is stronger than the previously known one.

In the quantum setting, we define the notion of quantum help and show in a more direct way that help and interaction are again equivalent. Moreover, we define quantum Non-Interactive Zero-Knowledge with classical help and prove that
it is equal to the class of languages that have classical honest-Verifier Zero Knowledge protocols secure against quantum Verifiers (\cite{Wat06, HKSZ07}). Last, we provide new complete problems for all these quantum classes.

Similar results were independently discovered by Dragos Florin Ciocan and Salil Vadhan.

\end{abstract}

\newpage

\setcounter{page}{1}

\section{Introduction}

In the setting of Zero-Knowledge, the Prover can prove to the
Verifier that the answer to an instance of a problem, e.g. an $NP$
problem with a witness $w$, is Yes without giving any other
information. In particular, the person that receives the proof
does not learn anything about $w$ or any other witness. In order to
create this kind of proofs, the Prover and the Verifier interact
with each other. The condition "without giving any other
information" has been formalized in \cite{GMR89,GMW91} and this security condition has been defined in the computational and the information-theoretic
setting.

We are interested in the information-theoretic setting and the class
$SZK$ (Statistical Zero-Knowledge) where an exponentially small
amount of information is leaked. This class has been widely studied
and many properties thereof are known (eg. \cite{Oka96,Vad99}). Some non-interactive models
have also been defined where there is a single message from the
Prover to the Verifier. If the Prover and Verifier do not share
anything in the beginning of the protocol, then the resulting class
is no larger than $BPP$. However, we can enhance the model, either
by having the Prover and Verifier share a uniformly random string
(the $NISZK$ class, see \cite{DMP88}, \cite{GSV99}) or some limited trusted help (the $NISZK_{|h}$
class).

The class $NISZK_{|h}$ was introduced by Ben-or and Gutfreund
\cite{GB00}. In this setting, the Prover and Verifier receive in the
beginning of the protocol some help from a trusted third party,
the {\em Dealer}. The Dealer has polynomial power, hence the help is
"limited", however he knows the input to the problem. They showed
that help does not add anything if we allow interaction
($SZK=SZK_{|h}$). They also described a complete problem for the
class $NISZK_{|h}$, the Image Intersection Density $(IID)$, and
showed that $NISZK \subseteq NISZK_{|h}  \subseteq SZK$, in other
words that help can always be replaced by interaction. They also
claimed to prove the opposite inclusion, $SZK \subseteq NISZK_{|h}$,
however they later retracted from this claim (\cite{BG03}).

In this paper, we start by proving that indeed help and interaction
are equivalent in Zero-Knowledge proofs, i.e. $SZK =
NISZK_{|h}$ (Section \ref{equivhi}). Our result can be thought of as showing that the power of $SZK$ lies only in the fact that there is a trusted access to the input (from the Verifier or from the Dealer). It will hopefully provide some more insight into the relation between the classes
$NISZK$ and $SZK$, which is a main open question in the area.
Moreover, we show that the $IID$
problem remains complete for a wider range of parameters. For
the proof we use a polarization lemma that is based on new
bounds on the Statistical Difference problem (Appendix \ref{polarization}).

In $2002$, Watrous defined a quantum analog of Zero-Knowledge proofs
(\cite{Wat02}) and studied the quantum class $QSZK$. Since then,
there has been a series of works that deal with the power and
limitations of quantum Zero-Knowledge proofs
(\cite{Kob03,Wat06,Kob07}) as well as attempts to find classical
interactive protocols that remain zero-knowledge even against
quantum adversaries (\cite{Wat06,HKSZ07}).

In the second part of our paper, we start by studying the class
$QNISZK$ that was defined by Kobayashi in \cite{Kob03}. Using new
results from \cite{BT07}, we give two complete problems for this
class, the Quantum Entropy Approximation ($QEA$) and the Quantum
Statistical Closeness to Uniform ($QSCU$). These complete problems
are the quantum equivalents of the complete problems for $NISZK$.
However, due to the fact that quantum expanders are different than
classical ones, the proof is different than in the classical case
(Section \ref{QNISZK}).

In addition, we study the role of help in quantum Zero-Knowledge
protocols. We define the notion of quantum help and show in a
straightforward way that it is again the case that help and
interaction are equivalent. We also define quantum Zero-Knowledge
with classical help, provide a complete problem for the class and
deduce that the message of the Prover can also be classical. This
allows us to prove that this class is equivalent to the class of
languages that have classical interactive protocols that remain
zero-knowledge even against quantum honest Verifiers (Section
\ref{Qhelp}).


\section{Preliminaries}

We start by describing some operations on probability distributions and proceed to provide definitions for classical and quantum Zero Knowledge classes and their complete problems.

\subsection{Operations on Probability distributions}

Let $X: \{0,1\}^{n} \rightarrow \{0,1\}^{m}$ be a polynomial size
circuit. The distribution encoded by $X$ is the distribution induced
on $\01^m$ by evaluating $X$ on a uniformly random input
from $\01^n$. We abuse notation and denote this distribution by $X$,
in other words, $X$ is both a circuit that encodes a distribution
and the distribution itself. Also, $\mathcal{P}_{n}$ is the set
of probability distributions on $\01^n$.

Denote by $SD(X,Y)$ the Statistical Difference between $X$ and $Y$, $SC(X,Y)$
their Statistical Closeness, $Disj(X,Y)$ the Disjointness of $X$
according to $Y$ and $\textrm{mut-}Disj$ the mutual Disjointness between $X$ and $Y$.
\begin{itemize}
\item $SD(X,Y) = \frac{1}{2} \sum_{i}|x_{i} - y_{i}| = 1 -  \sum_{i}\min(x_{i},y_{i})$
\item $SC(X,Y) = 1 - SD(X,Y) = \sum_{i}\min(x_{i},y_{i})$
\item $Disj(X,Y) = \frac{1}{2^n}| \{i \in \{0,1\}^{n} \ | \ \forall j \in
\{0,1\}^{n}, \ X(i) \neq Y(j) \}$
\item $\textrm{mut -}Disj(X,Y) = min(Disj(X,Y),Disj(Y,X))$
\end{itemize}

Note that $Disj(X,Y) \le SD(X,Y)$ and that $Disj(X,Y) \neq
Disj(Y,X)$ but mut-$Disj(X,Y) =
$mut-$Disj(Y,X)$.

\paragraph{Tensor Product}
$X \otimes Y$ corresponds to the distribution $(X,Y)$. If $X \in
\mathcal{P}_{n}$ and $Y \in \mathcal{P}_{m}$ then $X \otimes Y \in
\mathcal{P}_{n+m}$. We denote $X^{\otimes k}$ the distribution that
results by tensoring $X$ $k$ times.

\begin{prop}[Direct Product Lemmas]
Let $X,Y$ any probability distributions. Then,
\begin{enumerate}
\item $SD(X,Y) = \delta \implies  1 - 2\exp^{-k\delta^{2}/2} \le SD(X^{\otimes k},Y^{\otimes k}) \le k \delta$
\item $Disj(X,Y) = \delta \implies Disj(X^{\otimes k},Y^{\otimes k}) = 1 - (1 - \delta)^{k}$
\end{enumerate}
\end{prop}
\paragraph{XORing Distributions}
We define the $XOR$ operator which acts on a pair of distributions
and returns a pair of distributions. Let $(A,B) = XOR(X_{0},X_{1})$.
Then,
\[
\begin{array}{r}
A : \mathrm{pick } \ b \in_{R} \{0,1\}, \mathrm{return \ a \ sample \ of}\ X_{b} \otimes X_{b} \\
B : \mathrm{pick } \ b \in_{R} \{0,1\}, \mathrm{ return \ a \ sample
\ of }\ X_{b} \otimes X_{\bar{b}}
\end{array}
\]
\begin{prop}[XOR Lemmas] Let $X,Y$ probability distributions and $(A,B) =
XOR(X,Y)$. Then,
\begin{enumerate}
\item $SD(X,Y) = \delta \implies SD(A,B) = \delta^{2}$
\item mut-$Disj(X,Y) = \delta \implies $mut-$Disj(A,B) = \delta^{2}$
\end{enumerate}
\end{prop}

\paragraph{Flat Distributions} Let $X$ a distribution with
entropy $H(X)$. Elements $x_{i}$ of $X$ such that $|\log(x_{i}) +
H(X)| \le k$ are called $k$-typical. We say that $X$ is
$\Delta$-flat if for every $t>0$ the probability that an element
chosen from $X$ is $t\cdot \Delta$-typical is at least $1 -
2^{-t^{2} + 1}.$

\begin{prop}[Flattening Lemma] Let $X:\01^n \rightarrow \01^m$ a circuit that encodes a distribution. Then $X^{\otimes k}$ is $\sqrt{k}\cdot n$-flat.
\end{prop}

\paragraph{2-Universal hashing functions} A family $\mathcal{H}$ of
2-Universal hashing functions from $A \rightarrow B$ is such that
for every two elements $x,y \in A$ and $a,b \in B$
$
Pr_{h \in_{R} \mathcal{H}}[\ h(x) = a \ \textrm{and} \ h(y) = b] =
\frac{1}{|B|^{2}}.
$
\begin{prop}[Leftover hash lemma]
Let $\mathcal{H}$ a samplable family of 2-Universal hashing
functions from $A \rightarrow B$. Suppose $X$ is a distribution on
$A$ such that with probability at least $1-\delta$ over $x$ selected
from $X$, $Pr[X=x]\le \epsilon/|B|$. Consider the following
distribution
\begin{center}
$Z :$ choose $h \leftarrow \mathcal{H}$ and $x \leftarrow X$. return
$(h,h(x))$
\end{center}
Then, $SD(Z,I) \le O(\delta + \epsilon^{1/3})$, where $I$ is the
Uniform distribution on $\mathcal{H} \times B$.
\end{prop}

\subsection{Classical Zero Knowledge}

Zero Knowledge proofs are a special case of interactive proofs.
Here, we also want that the Verifier learns nothing from the
interaction other than the fact that $x \in \Pi_{Y}$ when it is the
case. The way it is formalized is that for $x \in \Pi_{Y}$, the
Verifier can simulate his view of the protocol defined by all the messages sent during the protocol as well as the verifier's private coins.
\begin{defi}
$\Pi \in SZK$ iff there exists an interactive protocol $\langle P,V
\rangle$ that solves $\Pi$ such that there exists a function $S$
computable in polynomial time and a function $\mu \in negl(k) \ll
1/poly(k)$ that has the following property :
\[
\forall x \in \Pi_{Y}, \ SD\left(S(x,1^{k}),\langle P,V
\rangle_{V}\right) \le \mu(k)
\]
\end{defi}
$S$ is called the simulator. We also have the following
non-interactive variants of $SZK$:

\noindent $\bullet$ {$\mathbf{NISZK}$ :} We suppose here that the
Prover and the Verifier additionally share a truly random string
$r$. We want the Verifier to be able to simulate both the random
string and the message $m_{P}$ from the Prover on Yes instances.

\begin{defi}
$\Pi \in NISZK$ iff with a truly random shared string $r$, there
exists an non-interactive protocol $\langle P,V \rangle$ that solves
$\Pi$ such that there exists a function $S$ computable in polynomial
time and a function $\mu \in negl(k) \ll 1/poly(k)$ that has the
following property :

\[
\forall x \in \Pi_{Y}, \ SD\left(S(x,1^{k}),(r,m_{P}(r,x))\right)
\le \mu(k)
\]
\end{defi}
\noindent
$\bullet$ {$\mathbf{NISZK_{|h}}$ :} We suppose here that the Prover
and the Verifier additionally share a string $h$ that is generated
by a trusted third party (the dealer) using some coins unknown to the verifier and the prover. This string is called the help and can
depend on the input. We want the Verifier to be able to simulate
both the help and the Prover's message on Yes instances.

\begin{defi}
$\Pi \in NISZK_{|h}$ iff there exists a
non-interactive protocol $\langle D,P,V \rangle$ that solves $\Pi$ where :
\begin{itemize}
\item The prover and the verifier share some help $h$ which is a random sample of $D$ depending on the input.
\item There exists a function $S$ computable in polynomial time
and a function $\mu \in negl(k) \ll 1/poly(k)$ that has the
following property :

\[
\forall x \in \Pi_{Y}, \ SD\left(S(x,1^{k}),(h,m_{P}(h,x))\right)
\le \mu(k)
\]
\end{itemize}
\end{defi}

\subsection{Quantum Statistical Zero Knowledge}

Quantum Statistical Zero Knowledge proofs are a special case of
Quantum Interactive Proofs. We can think of a quantum interactive
protocol $\langle P,V \rangle(x)$ as a circuit
$\left(V_{1}(x),P_{1}(x),\dots,V_{k}(x),P_{k}(x)\right)$ acting on
$\mathcal{V} \otimes \mathcal{M} \otimes \mathcal{P}$. $\mathcal{V}$
are the Verifier's private qubits, $\mathcal{M}$ are the message
qubits and $\mathcal{P}$ are the Prover's private qubits. $V_{i}(x)$
(resp. $P_{i}(x)$)  represents the $i^{th}$ action of the Verifier
(resp. the Prover) during the protocol and acts on $\mathcal{V}
\otimes \mathcal{M}$ (resp. $\mathcal{M} \otimes \mathcal{P}$).
$\beta_{i}$ corresponds to the state that appears after the $i^{th}$
action of the protocol.

In the Zero-Knowledge setting, we also want that the Verifier learns
nothing from the interaction other than the fact that $x \in
\Pi_{Y}$ when it is the case. The way it is formalized is that for $x
\in \Pi_{Y}$, the Verifier can simulate his view of the protocol.
We are interested only in protocols where the Verifier and the Prover use
unitary operations.

Let $\langle P,V \rangle$ a quantum protocol and $\beta_{j}$ defined
as before. The Verifier's view of the protocol is his private qubits
and the message qubits. $view_{\langle P,V \rangle}(j) =
Tr_{\mathcal{P}}(\beta_{j})$. We also want to separate the
Verifier's view whether the last action was made by the Verifier or
the Prover. We note $\rho_{0}$ the input state, $\rho_{i}$ the
Verifier's view of the protocol after $P_{i}$ and $\xi_{i}$ the
Verifier's view of the protocol after $V_{i}$.

We say that the Verifier's view can be simulated if on an input $x$,
there is a negligible function $\mu$ such that $\forall j$ we can
create $\sigma_{j}$ with quantum polynomial computational power such
that
\[
\| \sigma_{j} - view_{V,P}(j) \| \le \mu(|x|)
\]

Note that for a state $\sigma$ such that $\| \sigma - \rho_{i} \|
\le \mu(|x|)$ it is easy to see that $\sigma' = V_{i+1} \sigma
V^{\dagger}_{i+1}$ is close to $\xi_{i+1} = V_{i+1} \rho_{i}
V^{\dagger}_{i+1}$ in this sense that $\| \sigma' - \xi_{i+1} \| \le
\mu(|x|)$. Therefore we just need to simulate the $\rho_{i}$'s.

\begin{defi}
A protocol $\langle P,V \rangle$ has the zero-knowledge property for
$\Pi$ if for each input $x \in \Pi_{Y}$,  there is a negligible
function $\mu$ such that $\forall j$ we can create $\sigma_{j}$ with
quantum polynomial computational power such that
\[
\| \sigma_{j} - \rho_{j} \| \le \mu(|x|)
\]
\end{defi}

This formalizes the fact that on Yes instances, the Verifier does not
learn anything from the protocol except the fact that the input is a
Yes instance.

\begin{defi}
$\Pi \in QSZK$ iff there exists a quantum protocol $\langle P,V
\rangle$ that solves $\Pi$ and that has the zero-knowledge property
for $\Pi$.
\end{defi}

In the setting of Quantum Non-Interactive Statistical
Zero-Knowledge, first defined by Kobayashi \cite{Kob03}, the Prover
and Verifier share a maximally entangled state $\sum_i
\ket{i}\ket{i}$ and then the Prover sends a single quantum message
to the Verifier.

\begin{defi}
$\Pi \in QNISZK$ iff, when the Prover and Verifier share the maximally entangled state $\sum_i \ket{i}\ket{i}$, there exists a quantum non-interactive protocol $\langle P,V
\rangle$ that solves $\Pi$ and that has the zero-knowledge property
for $\Pi$.
\end{defi}

The notion of quantum help is more intricate and will be the subject of
Section \ref{Qhelp}.

\subsection{Complete problems for Zero-Knowledge classes}

The complete problems for the Zero-Knowledge classes are promise
problems. A promise problem $\Pi$ is defined by two disjoint sets
$\Pi_{Y}$ and $\Pi_{N}$. An instance $X$ of $\Pi$ is an
element of $\Pi_{Y} \cup \Pi_{N}$. We say that $\Pi$ reduces
to $\Omega$ ($\Pi \preccurlyeq \Omega$) iff there exists a poly-time
computable function $f$ such that
\[
X \in \Pi_{Y} \ \Rightarrow \ f(X) \in \Omega_{Y} \mbox{ and } X \in
\Pi_{N} \ \Rightarrow \ f(X) \in \Omega_{N}
\]
If $\Pi \preccurlyeq \Omega$ then $\Pi$ is no-harder than $\Omega$.
We can define the complement problem
$\overline{\Pi}$ as follows : $\overline{\Pi}_{Y} = \Pi_{N}$ and
$\overline{\Pi}_{N} = \Pi_{Y}$. In what follows, $X,Y$ are circuits
encoding probability distributions. \\

$SZK$-complete problems (see \cite{SV00}, \cite{GV98}) : \\

$\qquad$ \begin{tabular}{ll}

\textbf{{\em Statistical Difference} (SD)} & \textbf{ {\em Entropy Difference} (ED)}  \\ \\

$(X,Y) \in SD_{Y} \Rightarrow SD(X,Y) \ge 2/3$ & \quad $(X,Y) \in ED_{Y} \Rightarrow H(X) - H(Y) \ge 1$ \\
$(X,Y) \in SD_{N} \Rightarrow SD(X,Y) \le 1/3$ & \quad $(X,Y) \in ED_{N} \Rightarrow H(Y) - H(X) \ge 1$
\end{tabular} $ \ $ \\ \\

$NISZK$-complete problems (see \cite{GSV99}) : \\

$\qquad$\begin{tabular}{ll}

\textbf{{\em Entropy Approximation} (EA$^{t}$)} & \textbf{{\em Statistical Closeness to Uniform} (SCU)}  \\ \\

$ X \in EA^{t}_{Y} \Rightarrow H(X) \ge t + 1 $ & $ X \in SCU_{Y} \Rightarrow SD(X,I) \le 1/n $ \\
$ X \in EA^{t}_{N} \Rightarrow H(X) \le t - 1 $ & $ X \in SCU_{N} \Rightarrow SD(X,I) \ge 1 - 1/n $
\end{tabular} $ \ $  \\ \\

$NISZK_{|h}$-complete problem (see \cite{BG03}) : \\

$\qquad$\begin{tabular}{ll}
 \textbf{{\em Image Intersection Density} (IID)} & $ \textbf{{\em Mutual Image Intersection Density} (mut-IID)} $ \\ \\

$(X,Y) \in IID_{Y} \Rightarrow SD(X,Y) \le 1/n^2$ & $ (X,Y) \in $ mut-$IID_{Y} \Rightarrow SD(X,Y) \le 1/n^2 $ \\
$(X,Y) \in IID_{N} \Rightarrow Disj(X,Y) \ge 1 - 1/n^2$ & $ (X,Y) \in $ mut-$IID_{N} \Rightarrow \textrm {mut-}Disj(X,Y) \ge 1 - 1/n^2 $ \\
\end{tabular} $ \ $ \\ \\

Note that we can change the parameters to other parameters $\alpha$ and $\beta$. For example, $SD^{\alpha,\beta}$ corresponds to : $(X,Y) \in SD^{\alpha,\beta}_{Y} \implies SD(X,Y) \ge \alpha$ and $(X,Y) \in SD^{\alpha,\beta}_{N} \implies SD(X,Y) \le \beta$

Similarly, we can define the quantum equivalent problems $QSD$,
$QED$, $QEA^{t}$ and $QSCU$. In this case, $X,Y$ are the density
matrices that correspond to the output qubits of the circuits,
$SD(X,Y)$ is the trace distance and the entropy is the von Neumann
entropy.


\section{A new polarization lemma for the $IID$
problem}\label{pol}

The Zero-Knowledge protocols usually require from the promise
problems some parameters that are exponentially close to $0$ or $1$.
Polarizations are reductions from promise problems with worse
parameters to promise problems that can be solved by the protocol.
For example, there is a polarization for the $SD$ problem which
transforms $SD^{a,b}$ with $a^{2} > b$ to $SD^{1 - 2^{-k},2^{-k}}$
for any $k \in poly(n)$.

The best polarization that was known for $IID$ was that
$IID^{1/n^{2},1 - 1/n^{2}}$ reduces to $ IID^{2^{-k},1 - 2^{-k}}$
and henceforth $IID^{1/n^{2},1 - 1/n^{2}}$ is complete for
$NISZK_{|h}$ (\cite{BG03}). We will show here that $IID^{a,b}$ is complete for
$NISZK_{|h}$ with $ b > 2a $ ($a$ and $b$ are constants). We first improve an
upper bound on statistical difference and then use it to prove this
new polarization lemma for the $IID$ problem. The proofs are
presented in Appendix \ref{polarization}.

To prove a polarization lemma on the $SD$ problem, the following
bounds were used :
\begin{fact}[\cite{Vad99}] \label{old-ub}
Let $X,Y$ two probability distributions st. $SD(X,Y) = \delta$. Then
\[
1 - 2 \exp^{- k \delta^{2}/2} \le SD(X^{\otimes k},Y^{\otimes k})
\le k\delta
\]
\end{fact}
\noindent
We can improve the upper bound on Statistical Difference to
\[
SD(X^{\otimes k},Y^{\otimes k})  \le 1 - (1 - \delta)^{k} \le
k\delta
\]
by using the following lemma (proof in Appendix \ref{polarization}).
\begin{lemma} \label{new-ub}
Let $X,Y,Z,T$ four probability distributions with $SD(X,Y) =
\delta_{1}$ and $SD(Z,T) = \delta_{2}$. Then,
\[
SD(X \otimes Z, Y \otimes T) \le 1 - (1-\delta_{1})(1-\delta_{2}) =
\delta_{1} + \delta_{2} -\delta_{1}\delta_{2}
\]
\end{lemma}
\noindent
Using the new upper bound, we prove in Appendix \ref{polarization}
that

\begin{prop}\label{pola1}
mut-$IID^{a,b} \preccurlyeq $ mut-$IID^{2^{-k}, 1 - 2^{-k}}$ for $b > a$ ($a,b$ constants).
\end{prop}

To show completeness theorems and make the link with the $IID$ problem, we will use the following fact proven in \cite{BG03}.

\begin{fact}\label{f1}
Let $(X_{0},X_{1}) \in IID^{a,2b}$. Construct $(A,B)$ as following : \\ \\
A : pick $r \in_{R} \{0,1\}$ and $b \in_{R} \{0,1\}$, return $(X_{b}(r),b)$. \\
B : pick $r \in_{R} \{0,1\}$ and $b \in_{R} \{0,1\}$, return $(X_{b}(r),\overline{b})$ \\ \\
We have : $(X_{0},X_{1}) \in IID^{a,2b}_{Y} \Rightarrow (A,B) \in $ mut-$IID^{a,b}_{Y}$ and $(X_{0},X_{1}) \in IID^{a,2b}_{N} \Rightarrow (A,B) \in $ mut-$IID^{a,b}_{N}$. This being true when $b > a$.
\end{fact}

\begin{theo}\label{IID}
mut-$IID^{a,b}$ is complete for $NISZK_{|h}$ when $b > a$ and $IID^{a,b}$ is complete for
$NISZK_{|h}$ when $b > 2a$.
 $IID^{a,b}$ is $NISZK_{|h}$ complete for any
$a,b$ with $ b > 2a$ ($a,b$ constants).
\end{theo}

\begin{proof}
Let $a,b$ with $b > a$. Using the protocol of \cite{BG03}, we know that mut-$IID^{2^{-k}, 1 - 2^{-k}}$ is in $NISZK_{|h}$ so using Prop \ref{pola1},
mut-$IID^{a,b} \in NISZK_{|h}$. For hardness, we note that the proof in \cite{BG03} also extends when we replace $IID$ with mut-$IID$. mut-$IID \preccurlyeq $ mut-$IID^{a,b}$ (because $a,b$ are constants) and mut-$IID^{a,b}$ is hard for $NISZK_{|h}$. We extend this result by using the fact that $IID^{a,2b} \preccurlyeq $ mut-$IID^{a,b}$ (for $b > a$) using fact \ref{f1}.
\end{proof}

In the next section ,we will use this polarization lemma to show that $NISZK_{|h} = SZK$. This will, in turn, imply that $IID^{a,b}$ is complete for $b^2 > a$ using the polarization used for the $SD$ problem. Our initial polarization is still interesting because it shows that problems like $IID^{1/10,3/10}$ are in $SZK$, something which was not known before.

\section{Equivalence of help and interaction in Statistical Zero-Knowledge}\label{equivhi}

We show here that help and interaction are equivalent in the
Statistical Zero-Knowledge setting

\begin{theo}\label{equiv}
$SZK =  NISZK_{|h}$
\end{theo}

\begin{proof}
We know that $NISZK_{|h} \subseteq SZK$ because $IID$, the complete
problem of $NISZK_{|h}$, trivially reduces to $\overline{SD}$, the
complete problem of $SZK$. In what follows we also prove the
opposite inclusion, {\em i.e.} $SZK \subseteq NISZK_{|h} $ (Lemma
\ref{intertohelp}).
\end{proof}
In \cite{GB00}, the authors claimed to have proven this theorem, but
due to a flaw they retracted it in \cite{BG03}. Their
reduction from the $SZK$-complete problem $ED$ to $IID$ was in fact
only a reduction to $\overline{SD}$. Nevertheless, inspired by their
method we show a reduction from $\overline{EA}$ to $IID$.

In order to prove that help can replace interaction we follow \cite{GSV99} and
reduce the $SZK$-complete problem $\overline{ED}$ to several
instances of $EA$ and $\overline{EA}$ using the following fact :

\begin{fact}[\cite{GSV99}]

Let $X' = X^{\otimes 3}$ and $Y' = Y^{\otimes 3}$. Let $n$ the
output size of $X'$ and $Y'$. It holds that.
\begin{displaymath}
\begin{array}{l}
(X,Y) \in \overline{ED}_{Y} \Leftrightarrow \forall t \in \{1,\dots,n\}
\left[(X' \in {\overline{EA}}^{t}_{Y}) \vee (Y' \in
{EA}^{t}_{Y})\right] \\
(X,Y) \in \overline{ED}_{N} \Leftrightarrow \exists t \in \{1,\dots,n\}
\left[(X' \in \overline{EA}^{t}_{N}) \wedge (Y' \in {EA}^{t}_{N})
\right]
\end{array}
\end{displaymath}
\end{fact}

We know that $EA \in
NISZK_{|h}$ (since by definition $NISZK \subseteq NISZK_{|h}$) so it
remains to show the following two things:
\begin{enumerate}
\item $\overline{EA} \in NISZK_{h}$ : In order to this, we use similar tools to the ones in \cite{Vad99} and especially the "Complementary use of messages" originally used in \cite{Oka96}.
\item $NISZK_{|h}$ has some boolean closure properties : this will
allow us to reduce $\overline{ED}$ to a single instance of $IID$. In order to this, we use similar techniques than the ones used in \cite{SV98} since $\textrm{mut-}IID$ and $\overline{SD}$ are very similar.
\end{enumerate}

Note that this approach is similar to \cite{GSV99} in their attempt to show that $NISZK = SZK$. They showed that if $NISZK = co-NISZK$ then $NISZK = SZK$. We show here that $co-NISZK \subseteq NISZK_{|h}$ which suffices us (with the closure properties) to show that $NISZK_{|h} = SZK$.

\subsection{$\overline{EA}$ belongs to Non-Interactive Statistical Zero-Knowledge with help} $ \ $

To show that $\overline{EA} \in NISZK_{|h}$, we reduce the
$\overline{EA}$ problem to the $IID$ problem which is complete for
$NISZK_{|h}$.

Let $X$ an instance of $\overline{EA}^{t}$, {\em i.e.} an instance of
$\overline{EA}$ with approximation parameter $t$. Let $k = poly(m)$,
where $m$ is the input size and define $X' = X^{\otimes s}$ with $s = 4km^{2}$. Note that the input size of $X'$ is $m'=sm$ and $H(X') = sH(X)$. We have

\begin{claim}
Let $Z = X' \otimes I$, where $I$ is the uniform distribution. We can
create $Z'$ in polynomial time such that :
\begin{itemize}
\item $X \in \overline{EA}^{t}_{Y} \Rightarrow SD(Z,Z') \le 2^{- \Omega(k)}$
\item $X \in \overline{EA}^{t}_{N} \Rightarrow Disj(Z,Z') \ge 1 - 2^{- \Omega(k)}$
\end{itemize}
\end{claim}

\begin{proof}
Construct $Z'$ as following:
\\ \\
$Z'$ : choose $r \in_R \{0,1\}^{m'}, \ x = X'(r), \ h \in_R
\mathcal{H}_{m' + st,m'}, \ u \in_R \{0,1\}^{st}$. return
$(x,(h,h(r,u)))$. \\

Note that $Z'$ is of the form $Z' = X' \otimes A$ so we need to show
that, when fixing $x \in X'$, we have either $SD(I,A)$ small (in the
Yes instance) or $Disj(I,A)$ large (in the No instance). From the
Flattening lemma (see Preliminaries) we have

\begin{fact}\label{typical}$ \ $
\begin{enumerate}
\item $X'$ is $\Delta$-flat with $\Delta = 2\sqrt{k}m^{2}$. $s$ was
chosen such that $s = 2\sqrt{k}\Delta$.
\item Let $x \leftarrow X'$. $Pr[x$ is $\sqrt{k}\Delta$-typical$] \ge
1 - 2^{- \Omega(k)}$.
\end{enumerate}
\end{fact}

For $x \in X'$, let $wt(x) = \log |\{r \ | \ X'(r) = x\}|$. When $x
\in X'$ is fixed, the number of different possible inputs $(r,u)$
that are hashed is $2^{wt(x) + st}$. From the flattening lemmas, it
is easy to see that if $H(X) \le t-1$ then $wt(x)$ will be large
with high probability whereas if $H(X) \ge t + 1$ then $wt(x)$ will
be small with high probability. In more detail,

\begin{itemize}
\item[(i)] $\mathbf{H(X) \le t - 1}$.

For all $x \in X'$ which are $\sqrt{k}\Delta$-typical we have
$ \left| \log \frac{1}{2^{m'}} | \{r \ | \ X'(r) = x\}| + H(X') \right| \leq  \sqrt{k}\Delta$. Hence,
\[
wt(x)  \ge  {m' - sH(X) - \sqrt{k}\Delta} \ge {m' - st + s - \sqrt{k}\Delta} \ge {m' - st + \sqrt{k}\Delta}.
\]
Therefore, the number of inputs $(r,u)$ such
that $X'(r) = x$ and $u \in \{0,1\}^{st}$ is greater than $2^{m' +
\sqrt{k}\Delta} \ge 2^{m' + k}$. By the leftover hash lemma
(see Preliminaries), $SD((h,h(r,u)),I) \le O(2^{-
\Omega(k)})$. By Fact \ref{typical}, the probability of a
$\sqrt{k}\Delta$-typical $x$ is larger than  $\ge 1 - 2^{-
\Omega(k)}$ and hence we can conclude that $SD(Z,Z') \le 2^{-
\Omega(k)}$.

\item[(ii)] $\mathbf{H(X) \ge t + 1}$.

For all $x \in X'$ which are $\sqrt{k}\Delta$-typical we have
$$wt(x) \le {m' -
sH(X) + \sqrt{k}\Delta} \le {m' - st - s + \sqrt{k}\Delta} \le {m' -
st - \sqrt{k}\Delta}.$$

Therefore, the number of inputs $(r,u)$ such that $X'(r) = x$ and $u
\in \{0,1\}^{st}$ is smaller than $2^{m' - \sqrt{k}\Delta} \le 2^{m'
- k}$. Since we hash at most $2^{m' - k}$ values into
$\{0,1\}^{m'}$, we get only a $2^{-k}$ fraction of the total support
and hence $Disj(I,h(r,u)) \ge 1 - 2^{- \Omega(k)}$. By Fact
\ref{typical}, the probability of a $\sqrt{k}\Delta$-typical $x$ is
larger than  $\ge 1 - 2^{- \Omega(k)}$ and hence we can conclude
that $Disj(Z,Z') \geq 1 - 2^{- \Omega(k)}$.
\end{itemize} \end{proof}
From the distribution $X$, we have created $Z,Z'$ in polynomial time such that :
\begin{itemize}
\item $X \in \overline{EA}_{Y} \Rightarrow (Z,Z') \in IID_{Y}$.
\item $X \in \overline{EA}_{N} \Rightarrow (Z,Z') \in IID_{N}$.
\end{itemize}
So $\overline{EA} \preccurlyeq IID$ and from the completeness of
$IID$ for $NISZK_{|h}$, we have $\overline{EA} \in NISZK_{|h}$.

\subsection{Closure properties for $NISZK_{|h}$}

Closure properties have been widely used in the study of Zero-Knowledge classes (see \cite{DDPY94} or \cite{SV98}).
Every promise problem $\Pi \in NISZK_{|h}$ reduces to the $\textrm{mut -}IID$
promise problem and hence, we just have to concentrate on this problem. Note
that this problem is very similar to the $\overline{SD}$ promise
problem and hence we use similar techniques to those used to show closure
properties for $SZK$ from the $SD$ problem. In our case, we just
need to show some limited closure properties that will be enough to
prove that $\overline{ED} \in NISZK_{|h}$.

\begin{defi} Let $\Pi^{1},\dots,\Pi^{k} $ some promise problems. We define
$AND(\Pi^{1},\dots,\Pi^{k}) :$
\begin{itemize}
\item $ (X^{1}, \dots ,X^{k}) \in AND(\Pi^{1},\dots,\Pi^{k})_{Y}
\Rightarrow \forall i \in \{1,\dots,k\} \ X^{i} \in \Pi^{i}_{Y} $
\item $ (X^{1},\dots,X^{k}) \in AND(\Pi^{1},\dots,\Pi^{k})_{N}
\Rightarrow \exists i \in \{1,\dots,k\} \ X^{i} \in \Pi^{i}_{N} $
\end{itemize}
\end{defi}

In the $AND$ definition, we assume $k$ to be of size polynomial in the input size, i.e. $k \in poly(n)$.

\begin{defi} Let $\Pi,\Omega$ two promise problems. We define
$OR(\Pi,\Omega)$ :
\begin{itemize}
\item $ (X,Y) \in OR(\Pi,\Omega)_{Y} \Rightarrow X \in \Pi_{Y} \ or \
Y \in \Omega_{Y} $
\item $ (X,Y) \in OR(\Pi,\Omega)_{N} \Rightarrow X \in \Pi_{N} \ and
\ Y \in \Omega_{N} $
\end{itemize}
\end{defi}

We will show that $NISZK_{|h}$ is closed under $AND$ and $OR$ which
is enough for our purposes.

\begin{claim}
$NISZK_{|h}$ is closed under $AND$.
\end{claim}

\begin{proof}
Let $\Pi^{1},\dots,\Pi^{k}$ in $NISZK_{|h}$ and $(A^{1},\dots,A^{k})$ an instance
of $AND(\Pi^{1},\dots,\Pi^{k})$. We reduce each $\Pi^{i}$ to the
$\textrm{mut-}IID$ problem which means that we transform each $A^{i}$ into a pair of distributions $(X^{i},Y^{i})$ such that $A^{i} \in
\Pi^{i}_{Y} \Rightarrow (X^{i},Y^{i}) \in \textrm{mut-}IID_{Y}$ and $A^{i} \in
\Pi^{i}_{N} \Rightarrow (X^{i},Y^{i}) \in \textrm{mut-}IID_{N}$. Let $X = X^{1}
\otimes \dots \otimes X^{k}$ and $Y = Y^{1} \otimes \dots \otimes
Y^{k}$. We first polarize each pair $(X^{i},Y^{i})$ such that
$(X^{i},Y^{i}) \in \textrm{mut-}IID^{1/n^{2}k,1 - 1/n^{2}}$ (which is possible since $k \in poly(n)$). Then, we use the following fact from \cite{Vad99} and \cite{BG03}:
\begin{fact} $ \ $
\begin{itemize}
\item $SD(X,Y) \le \sum_{i} SD(X^{i},Y^{i})$
\item $\textrm{mut-}Disj(X,Y) \ge \max_{i} \textrm{mut-}Disj(X^{i},Y^{i})$
\end{itemize}
\end{fact}

From this fact, we can easily see that $(A^{1},\dots,A^{k}) \in
AND(\Pi^{1},\dots,\Pi^{k})_{Y} \Rightarrow (X,Y) \in \textrm{mut-}IID_{Y}$ and
that $(A^{1},\dots,A^{k} )\in AND(\Pi^{1},\dots,\Pi^{k})_{N}
\Rightarrow (X,Y) \in \textrm{mut-}IID_{N}$, which concludes our proof.
\end{proof}

\begin{claim}
$NISZK_{|h}$ is closed under $OR$.
\end{claim}

\begin{proof}
Let $\Pi,\Omega \in NISZK_{|h}$. Let $I$ an instance of $\Pi$ and $J$ an
instance of $\Omega$. We reduce $I$ to a pair of distributions
$(X'_{0},Y'_{0})$ such that $I \in \Pi_{Y} \Rightarrow (X'_{0},Y'_{0})
\in \textrm{mut-}IID_{Y}$ and $I \in \Pi_{N} \Rightarrow (X'_{0},Y'_{0}) \in
\textrm{mut-}IID_{N}$. Similarly, we reduce $J$ to a pair of distributions
$(X'_{1},Y'_{1})$. Using our polarization, we create
$(X_{0},Y_{0})$ and $(X_{1},Y_{1})$ that are instances of mut-$IID^{1/n^2,\sqrt{(1-1/n^2)}}$
Now, consider the following two distributions\\

A : pick $b \in_R \{0,1\}$, return a sample of $X_{b} \otimes
Y_{b}$.

B : pick $b \in_R \{0,1\}$, return a
sample of $X_{b} \otimes Y_{\bar{b}}$. \\

This is a generalization of the $XOR$ transformation and was used in
\cite{Vad99} to show closure properties for $SZK$.
We now use the following fact
\begin{fact}\cite{Vad99} and \cite{BG03}
\begin{itemize}
\item
$SD(A,B) = SD(X_{0},Y_{0}) * SD(X_{1},Y_{1})$
\item
$\textrm{mut-}Disj(A,B) = \textrm{mut-}Disj(X_{0},Y_{0}) * \textrm{mut-}Disj(X_{1},Y_{1})$
\end{itemize}
\end{fact}

From this, we can easily see that $(X_{0},Y_{0}) \in $ mut-$IID^{1/n^2,\sqrt{(1-1/n^2)}}_{Y}$ or
$(X_{1},Y_{1}) \in $ mut-$IID^{1/n^2,\sqrt{(1-1/n^2)}}_{Y} \Rightarrow (A,B) \in IID_{Y}$.
Similarly, if $(X_{0},Y_{0}) \in $ mut-$IID^{1/n^2,\sqrt{(1-1/n^2)}}_{N}$ and $(X_{1},Y_{1}) \in
$ mut-$IID^{1/n^2,\sqrt{(1-1/n^2)}}_{N} \Rightarrow (A,B) \in $ mut-$IID_{N}$. We have therefore reduced
$OR(\Pi,\Omega)$ to a single instance of mut-$IID$. Since mut-$IID$ is in $NISZK_{|h}$  we conclude that $OR(\Pi,\Omega) \in NISZK_{|h}$.
\end{proof}

\subsection{Help can replace interaction}
We can now prove that help can replace interaction and hence
conclude the proof of Theorem \ref{equiv}.
\begin{lemma}\label{intertohelp}
$SZK \subseteq NISZK_{|h} $
\end{lemma}

\begin{proof}
We show that $\overline{ED} \in NISZK_{h}$, which will allow us to
conclude since $\overline{ED}$ is complete for $SZK$. Let $(X,Y)$ an
instance of $\overline{ED}$.

We have already shown that $EA$ and $\overline{EA}$ are in $NISZK_{|h}$.
Moreover, we have closure under $OR$, and hence for all $t$ there
exists a promise problem $\Pi^{t} \in NISZK_{|h}$ and an input
$A^{t}$ such that
\[
\begin{array}{l}
(X',Y') \in OR(\overline{EA}^{t},EA^{t})_{Y} \Rightarrow A^{t} \in
\Pi^{t}_{Y} \\
(X',Y') \in OR(\overline{EA}^{t},EA^{t})_{N} \Rightarrow A^{t} \in
\Pi^{t}_{N}
\end{array}
\]
Therefore,
\begin{displaymath}
\begin{array}{l}
(X,Y) \in \overline{ED}_{Y} \Rightarrow \forall t \in \{1,\dots,n\}
A^{t} \in \Pi^{t}_{Y} \\
(X,Y) \in \overline{ED}_{N} \Rightarrow \exists t \in \{1,\dots,n\} \ A^{t} \in
\Pi^{t}_{N}
\end{array}
\end{displaymath} $ \ $ \\
and from the closure under $AND$ we conclude that $\overline{ED} \in
NISZK_{|h}$.
\end{proof}

This theorem has some interesting corollaries.
\begin{corol}
$NISZK_{|h}$ has all the properties of $SZK$ like closure under
complement or closure under boolean formula.
\end{corol}

It is interesting to find a non-interactive class that has all the
properties of $SZK$. It means that the power of $SZK$ lies only in
the fact that there is a trusted access to the distributions (from
the Verifier or from the Dealer).

\begin{corol}
The $IID$ problem is complete for $SZK$.
\end{corol}

We have here a new complete problem for $SZK$. This problem is
easier to manipulate and could be used to find other results about
$SZK$.


\section{Complete problems for $QNISZK$}\label{QNISZK}

In this section we study complete problems for the class $QNISZK$. Note that
Kobayashi showed a complete problem for the case of Non-Interactive
Perfect Zero-Knowledge, however was unable to extend his proof to
the case of Statistical Zero-Knowledge.

We continue this line of work and give two complete problems for
$QNISZK$, the {\em Quantum Entropy Approximation} and the {\em
Quantum Statistical Closeness to Uniform}. These are the natural
generalizations of the $NISZK$-complete problems $EA, SCU$. Ben-Aroya and Ta-Shma showed that $QEA$ reduced to $QSD$. In fact, during their proof, they showed that $QEA \in QSCU^{a,b}$ but these parameters $a,b$ were not good enough to show that $QEA \in QNISZK$. We will modify their proof to show that $QEA \in QNISZK$ and then conclude using similar techniques than the ones used in the classical case (see \cite{GSV99} as well as the analysis of $QNISZK$ done by Kobayashi \cite{Kob03}). The
proof will follow from the following three lemmas.

\begin{lemma}
$QEA \in QNISZK$.
\end{lemma}
\begin{proof}
We modify the proof of \cite{BT07} to show that $QEA \in QNISZK$.
Let $X$ an instance of $QEA^{t}$ with input size $m$ and
$\mathbb{I}$ the totally mixed state.

\begin{claim}[\cite{BT07}]
We can create $X'$ such that
\begin{itemize}
\item $X \in QEA_{Y} \Rightarrow SD(X',\mathbb{I}) \le 5 \epsilon$
\item $X \in QEA_{N} \Rightarrow SD(X',\mathbb{I}) \ge
\frac{1}{2qm}$
\end{itemize}
where $q \ge 2\log(1/ \epsilon) + \log(qm) + O(1)$ and also
$q \ge \sqrt{\log(1/ \epsilon)}\sqrt{q}n + 1$. \\
\end{claim}
We apply this claim with the following parameters : fix $\epsilon =
2^{-k}$ with $k \in poly(n)$ and then $q \in poly(n)$ that satisfies
the constraints. Let $X'$ be the resulting distribution. Now let $r
= 8k(qm)^{2} \in poly(n)$ and $Y = X'^{\otimes r}$. By using bounds
on Statistical Difference, we have
\begin{itemize}
\item $X \in QEA_{Y} \Rightarrow SD(X',\mathbb{I}) \le 5r \epsilon \le 2^{-\Omega(k)}$
\item $X \in QEA_{N} \Rightarrow SD(X',\mathbb{I}) \ge 1 - 2^{-k}$
\end{itemize}
Kobayashi showed in \cite{Kob03} that $QSCU^{2^{-k},1-2^{-k}} \in
QNISZK$ and hence by our claim that $QEA \preccurlyeq
QSCU^{2^{-k},1-2^{-k}}$ we conclude that $QEA \in QNISZK$.
\end{proof}

\begin{lemma}
$QSCU \preccurlyeq QEA$.
\end{lemma}

\begin{proof}
We use the following fact about the relation of trace distance and von
Neumann entropy
\begin{fact} Let X a quantum state of dimension $n$.
\begin{enumerate}
\item $\|X - \mathbb{I}\|_{tr} \le \alpha \Rightarrow S(X) \ge n(1 - \alpha -
1/2^{n})$.
\item $\|X - \mathbb{I}\|_{tr} \ge \beta \Rightarrow S(X) \le n -
\log(\frac{1}{1-\beta})$.
\end{enumerate}
\end{fact}
Let $X$ a quantum mixed state of dimension $n
\ge 16$. $\|X - \mathbb{I}\|_{tr} \le 1/n \Rightarrow S(X) \ge n -
2$. $\|X - \mathbb{I}\|_{tr} \ge 1 - 1/n \Rightarrow S(X) \le n -
4$. When $n \le 16$, we can solve $QSCU$ polynomially. We have a
reduction from $QSCU$ to $QEA$.
\end{proof}

\begin{lemma}
Every problem in $QNISZK$ reduces to $QSCU$.
\end{lemma}
\begin{proof}
The proof of hardness for $QNIPZK$ extends naturally to this
problem. We will not repeat the proof here. The interested reader
can see \cite{Kob03} for this proof.
\end{proof}
It now follows immediately that
\begin{theo}
$QEA$ and $QSCU$ are complete for $QNISZK$.
\end{theo}
\begin{proof} $QSCU$ is hard for $QNISZK$ and $QSCU \preccurlyeq QEA$ so both problems are hard for $QNISZK$. $QEA \in QNISZK$ and $QSCU \preccurlyeq QEA$
so they are both in $QNISZK$.
\end{proof}


\section{Help in quantum Non-Interactive Zero-Knowledge protocols}
\label{Qhelp}

In classical Non-Interactive Zero-Knowledge, the Prover and Verifier
start with a shared uniformly random string, which is independent of
their input. Classical help was a natural generalization of this and
was defined as a shared string created by a trusted third party with
polynomial power (the Dealer) who has access to the input.

In quantum Non-Interactive Zero-Knowledge, the Prover and Verifier
share a maximally entangled state $\sum_{i}|i\rangle|i\rangle$, with
the Prover having the first register and the Verifier the second.
Note that this state is pure and  independent of the input $x$.

\paragraph{Help with unitaries}

We define quantum help as a generalization of the maximally
entangled state. We suppose here that there is a trusted Dealer with
quantum polynomial power that performs a unitary $U_x$ and creates a
state $h_{PV}$ in the space $\mathcal{P} \times \mathcal{V}$. The
Prover gets $h_{P} = Tr_{\mathcal{V}}(h_{PV})$ and the Verifier gets
$h_{V} = Tr_{\mathcal{P}}(h_{PV})$. Note that the state $h_{PV}$ is
a pure state and depends on the input.

\begin{defi}
We say that $\Pi \in QNISZK_{|h}$ if there is a non-interactive
protocol $\langle D,P,V \rangle$ that solves $\Pi$ with the
Zero-Knowledge property, where the Verifier and the Prover share a
pure state $h_{PV}$ created by a Dealer $D$ that has quantum
polynomial power and access to the input. They also start with
qubits initialized at $|0\rangle$. We denote by $\langle D,P,V
\rangle$ the entire protocol.
\end{defi}
Next, we prove that help and interaction are equivalent in the
quantum setting, but with a much easier proof than in the classical
case.

\begin{theo}
$QNISZK_{|h} = QSZK$
\end{theo}

\begin{proof}
We start by showing that $QNISZK_{|h} \subseteq QSZK$. Let $\Pi \in
QNISZK_{|h}$ and $\langle D,P,V \rangle $ denote the protocol. Since
$h_{PV}$ is a pure state, we can create another protocol $\langle
\widetilde{P},\widetilde{V}\rangle$ where the Verifier takes the
place of the Dealer. Because the Dealer is a unitary (and has no
private qubits), this can be done. The protocol is the same so
soundness and completeness are preserved. The first message in
$\langle \widetilde{P},\widetilde{V}\rangle$ can be simulated
because the circuit of the Dealer is public and computable in
quantum polynomial time. The second message in $\langle
\widetilde{P},\widetilde{V}\rangle$ can be simulated because of the
Zero-Knowledge property of the protocol $\langle P,V \rangle$.

The inclusion  $QSZK \subseteq QNISZK_{|h} $ is immediate, since
there exists a two message protocol for a $QSZK$-complete problem
(see \cite{Wat02}). The first message of the Verifier can be
simulated by the Dealer's help.
\end{proof}

\paragraph{Using non-unitaries}
The unitary restriction is natural when dealing with quantum Zero-Knowledge classes. However, unitary help does not allow the dealer to keep some information private. In fact, we can imagine a stronger quantum help, where the Dealer can perform any quantum operation in order to create the help. For example, he can create a quantum state, keep part of it to himself and share the rest of the state between the Prover and the Verifier.

It is not hard to see, that in this way, the dealer can create an even stronger type of classical help, namely where he can give secret correlated messages to the Verifier and the Prover. Since we know that
$NISZK^{SEC} = AM$ (see \cite{PS05}) we can conclude that non-unitary help is very strong. Note also that with non-unitaries we don't know if help and interaction are equivalent. The case of Quantum Zero Knowledge protocols with non-unitary players is indeed very interesting and we refer the reader to
\cite{CK07} for more results.

\subsection{Quantum Non-Interactive Zero-Knowledge with classical help} \label{QChelp}

We now define two "hybrid" classes, where the Prover and
Verifier are quantum, however in the beginning of
the protocol they only share classical information. These classes
have very interesting connections to the class of languages that
possess classical zero-knowledge protocols secure against quantum
adversaries, {\em i.e.} the class studied by Watrous \cite{Wat06} and Hallgren {\em et al} \cite{HKSZ07}. We start by providing some appropriate definitions.

\begin{defi}
We say that a circuit $C$ is $\epsilon$-probabilistic if
\[
\forall x, \ \exists ! y, \ Pr(C(x) = y) \ge 1 - \epsilon
\]
This $y$ will be called the natural image of $x$ and will be noted $Nat_{C}(x)$
\end{defi}

We now define $q$-samplable distributions as follows:
\begin{defi}
A distribution $D \in \mathcal{P}$ is called $q$-samplable if
it can be
represented by a $2^{-k}$-probabilistic circuit $C$ ($k \in poly(n)$) with classical input and output and such that in order to compute $C(x)$ for any $x$, we need a $BQP$ machine.
\end{defi}

\COMMENT{Note that in the classical case, this problem can be dealt with because the only source of randomness come from coins that and these coins can be put in the input. However, this cannot be done in the quantum case because randomness also comes from measurements. Fortunately, we will be able to deal with circuits that have limited randomness. We will call these circuits $\epsilon$-probabilistic circuits and we define them as follows :
\begin{defi}
We say that a circuit $C$ is $\epsilon$-probabilistic if
\[
\forall x, \ \exists ! y, \ Pr(C(x) = y) \ge 1 - \epsilon
\]
This $y$ will be called the natural image of $x$ and will be noted $Nat_{C}(x)$
\end{defi}

}

To deal with $q$-samplable distributions, we also extend the definition of Disjointness to probabilistic circuits.

\begin{defi}
\[
Disj(X,Y) = \frac{1}{2^{n}} \sum_{r \in \{0,1\}^{n}}{ \max_{y}( Pr(Y(y) = X(r)))}
\]
$Disj(X,Y)$ must be understood as follows : "If I take a random $x$ of $X$, and I'm given a $y$ (potentially the best), what is the probability that $Y(y) = x$ ?"
\end{defi}

Note that when the second distribution ($Y$) is described by a deterministic circuit then this notion of disjointness is equivalent to the original one.

From this fact, we will show a simple relationship between Statistical Difference and Disjointness. In the case of deterministic distributions, we know that $Disj(X,Y) \le SD(X,Y)$.

\begin{lemma}\label{SDtoDisj}
Let $(X,Y)$ be $2$ $\epsilon$-probabilistic circuits. We have :
$Disj(X,Y) \le SD(X,Y) + 2\epsilon$.
\end{lemma}
\begin{proof}
Let $(X,Y)$ be $2$ $\epsilon$-probabilistic circuits. We define $\widetilde{Y}$ as following : $\widetilde{Y}(r) = Nat_{Y}(r)$. We can easily see that $SD(\widetilde{Y},Y) \le \epsilon$ and that $Disj(X,Y) \le Disj(X,\widetilde{Y}) + \epsilon$. From this, we conclude that :
\[
Disj(X,Y) \le Disj(X,\widetilde{Y}) + \epsilon \le SD({X},\widetilde{Y}) + \epsilon \le SD(X,Y) + 2\epsilon
\]
\end{proof}

Note that $2^{-n}$-probabilistic circuits behave similarly (with exponentially small difference) to deterministic circuits. This means that we can apply polarization lemmas and extend all the completeness theorems that were shown with classical distributions to $q$-samplable distributions. We can now study $QNISZK_{|ch}$.

\begin{defi}
We say that $\Pi \in QNISZK_{|ch}$ if there exists a non-interactive
protocol $\langle P,V \rangle$ that solves $\Pi$ with the
Zero-Knowledge property where the Verifier and the Prover start with
some classical help $h$ distributed over a distribution $D$ prepared
by a trusted Dealer with quantum polynomial power. We want the dealer $D$ and the simulation $S$ to be $q$-samplable distributions. The prover and the verifier also start
with $|0\rangle$ qubits.  We denote $\langle D,P,V\rangle$ the
entire protocol.
\end{defi}

Let us define the problem $IID^{q}$:
Let $X,Y$ two $q$-samplable probability distributions which are describes by $2^{-n}$-probabilistic circuits
\begin{itemize}
\item $(X,Y) \in IID^{q}_{Y} \Rightarrow SD(X,Y) \le 1/4$
\item $(X,Y) \in IID^{q}_{N} \Rightarrow Disj(X,Y) \ge 3/4$
\end{itemize}

We prove that this problem is complete for $QNISZK_{|ch}$ by the following two lemmas.

\begin{lemma}\label{ChMemb}
$IID^{q} \in QNISZK_{|ch}$.
\end{lemma}

\begin{proof}
Let $(X,Y)$ an instance of $IID^{q}$. Using our polarization lemma,
we construct $(X',Y')$ such that $(X,Y) \in IID^{q}_{Y} \Rightarrow SD(X',Y') \le 2^{-k}$ and $(X,Y) \in IID^{q}_{N} \Rightarrow Disj(X',Y') \ge 1 - 2^{-k}$ for some $k \in poly(n)$. We use the same protocol as for the classical case: \\

\cadre{
\begin{center}
\textbf{ Protocol in $QNISZK_{|ch}$ for the $IID^{q}$ problem }
\end{center}
H : create $x' \leftarrow X'$ and reveal it. \\ \\
P : send $r$ such that $Y'(r) = x'$. \\ \\
V : Verify that $Y'(r) = x'$ \\
} $ \ $ \\\\
This protocol is the same as the one used in \cite{BG03}.
Note that the completeness and soundness correspond exactly to the Disjointness of the two distributions and hence they follow from Lemma \ref{SDtoDisj}. Moreover, working on $q$-samplable distributions doesn't change the Zero-Knowledge property and hence it follows immedaitely from \cite{BG03}.
\end{proof}

\begin{lemma}\label{ChHard}
Every problem in $QNISZK_{|ch}$ reduces to $IID^{q}$
\end{lemma}

\begin{proof}
The proof of Ben-Or and Gutfreund that $IID$ is hard for $NISZK_{|h}$ can be naturally extended to the case where the Verifier and the Dealer are $BQP$ machines by taking into account that the distributions are now $q$-samplable.

Consider a promise problem $\Pi \in QNISZK_{|ch}$. Let $\langle D,P,V
\rangle$ be a non-interactive protocol for $\Pi$ with completeness $c(k)$, soundness $s(k)$and simulator deviation $\mu(k)$ with $1 - c(k),s(k),\mu(k) \in negl(k)$. Let $x$ an instance of $\Pi$. Consider now the
two following distributions :

$D_{0}$ : run the Dealer $D$ on $x$.

$D_{1} : $ run the simulator $k \in poly(n)$ times on $x$ with the same coins to get $k$ samples $(h,m_{P})$. Note that these copies are the same with exponentially high probability because the simulator is $2^{-O(k)}$-probabilistic. Run
the accepting procedure $A$ on each copy of $(x,h,m_{P})$. Output $h$ if $V$ accepts the majority of the times and $\bot$ otherwise.

\begin{itemize}

\item If $x \in \Pi_{Y}$ then the
Verifier will accept the majority of times with probability $(1 - 2^{-O(k)})$ because of completeness.
In this case, the distribution $D_{1}$ is equal to the
simulation of the help, which has statistical difference $\mu(k)$
from the real help. Since the distribution $D_{0}$ is the
distribution of the real help, we have $SD(D_{0},D_{1}) \le \mu(k) + 2^{-O(k)} \le 1/4$ and $(D_{0},D_{1}) \in IID^{q}_{Y}$.

\item Let $x \in \Pi_{N}$ and $B$ be the set of help strings, such that $h \in B \Rightarrow \exists \ m_P \ Pr[A(x,h,m_P) = Yes] \ge 1/3$ where $A$ is the verifying procedure of $V$. The
probability that $D_{0}$ produces a sample $h \in B$ (and therefore a sample in $B \cup \{\bot\}$) is $\le 3s(k)$ due to the soundness condition. It also holds that the probability that $D_1$ produces a sample in $B \cup \{\bot\}$
is $\ge 1 - O(2^{-k})$. This can seen as follows: the probability that $D_1$ outputs $h \in \overline{B}$ is equal to the probability that the Verifier accepts the majority of times, when running $A$ $k$ times with $h \in \overline{B}$, which happens with probability at most $2^{-O(k)}$.
We conclude that
$Disj(D_{0},D_{1}) \ge (1 - 2^{(O(k))})(1 - 3s(k)) \ge 3/4$ and
$(D_{0},D_{1}) \in IID^{q}_{N}$
\end{itemize}
Since the Dealer and Simulator are $q$-samplable, the
distributions $D_{0}$ and $D_{1}$ are also $q$-samplable.
\end{proof}

and hence $D_{1}$ is $2^{-O(k)}$-probabilistic

From Lemma \ref{ChMemb} and Lemma \ref{ChHard}, we have
\begin{theo}
 $IID^{q}$ is complete for $QNISZK_{|ch}$.
\end{theo}

Similarly, we can define Quantum Non-Interactive Zero-Knowledge
where the Prover and the Verifier share a classical random string.
We denote this class $QNISZK_{r}$. Let us define $SCU^{q}$ as the
statistical closeness to uniform applied on a $q$-samplable
distribution. By the same arguments $SCU^{q}$ is
complete for $QNISZK_{|r}$.

Using these complete problems, we have the following interesting
corollary
\begin{corol}\label{classical}
In $QNISZK_{|r}$ and $QNISZK_{|ch}$, the Prover sends a classical
message.
\end{corol}
\begin{proof}
This is true because there is a protocol for $IID^{q}$ and $SCU^{q}$
where the Prover sends a classical message and these two problems
are complete.
\end{proof}

Now denote by $SZK_{q}$ the class $SZK$ where the Verifier and
simulation use quantum polynomial power. In other words, this is the
class of languages that have classical protocols which are
Zero-Knowledge against quantum Verifiers. Similarly, define the
classes $HVSZK_q$ and $NISZK_{|h,q}$ (where both the Verifier and
the Dealer use quantum power). The class $SZK_q$ was studied by
Watrous (\cite{Wat06}) and Hallgren {\em et al} \cite{HKSZ07}. It
remains open to show whether these three classes are equal to each
other, which is true when the Verifier is classical.

Note that by corollary \ref{classical}, we have that $QNISZK_{|ch} =
NISZK_{|h,q}$. Using our analysis of $NISZK_{|h}$, we can show the
following :
\begin{theo}
$NISZK_{|h,q} = HVSZK_{q}$
\end{theo}
\begin{proof}
 Similar to the case of $HVSZK$, we can show that
$SD^{q}$ is complete for $HVSZK_{q}$ (see also \cite{Vad99}) where
$SD^{q}$ is the natural extension of $SD$ applied to $q$-samplable
distributions. From section \ref{equivhi}, we know a reduction from
$SD$ to $IID$. The same reduction works from $SD^{q}$ to $IID^{q}$
so $HVSZK_{q} \subset QNISZK_{|ch} = NISZK_{|h,q}$. Because
$IID^{q}$ trivially reduces to $SD^{q}$, we have $HVSZK_{q} =
NISZK_{|h,q}$.
\end{proof}

\section{Conclusion and further work}
Our work settles the question of the role of help in Zero-Knowledge protocols by showing that it is equivalent to interaction. In other words, we showed that the only thing that is important to create a statistical Zero-Knowledge proof is a trusted access to the input (from the Dealer or from the honest Verifier). This will hopefully shed some light into the relation of Non-Interactive and Interactive Zero-Knowledge, which still remains open.

In the quantum setting, we gave the first formal definition
of help for Zero-Knowledge protocols. We showed that quantum help is also equivalent to interaction and that the case of classical help is closely related to the class of languages that have classical zero-knowledge protocols secure against quantum Verifiers. It would be interesting
to see if quantum help could also give some interesting results
concerning the class $SZK_{q}$, and especially whether $SZK_{q} =
HVSZK_q$.


\bibliographystyle{alpha}

\newpage
\begin{appendix}

\section{Details of the polarization of $IID$}\label{polarization}

\begin{proof}(Lemma \ref{new-ub})
Define $w_{S}(X) = \sum_{i\in S}x_{i}$ to be the weight of $X \in
P_{n}$ on the set $S \subseteq \01^n$, $S(X,Y) = \{i \in \01^n | \
x_{i} \le y_{i} \}$  and $\overline{S}(X,Y)$ the complement. Fix
$X,Y,Z,T$ four probability distributions with $c_{1} = 1 -
\delta_{1} = SC(X,Y)$,  $c_{2} = 1 - \delta_{2} = SC(Z,T)$ and $c =
1 - \delta = SC(X \otimes Z, Y \otimes T)$. Let $A = S(X,Y)$, $A' =
S(Z,T)$, $\overline{A}$ and $\overline{A'}$ the complementary sets,
$\alpha_{1} = w_{A}(X)$, $\beta_{1} = w_{A}(Y)$, $\alpha_{2}
=w_{A'}(Z)$ and $\beta_{2} = w_{A'}(T)$. We have :
\[
c_{1} = \sum_{i}\min(x_{i},y_{i}) = w_{A}(X) + w_{\overline{A'}}(Y)
= \alpha_{1} + 1 - \beta_{1} \;\;\;\;\;\;\ \mbox{and} \;\;\;\;\;\;
c_{2} = \alpha_{2} + 1 - \beta_{2}
\]
We now show that $c \ge c_{1}c_{2}$.
\begin{eqnarray*}
c & = & \sum_{i,j}\min(x_{i}z_{j},y_{i}t_{j})  \\
& = & \sum_{i\in A , j\in A'} \min(x_{i}z_{j},y_{i}t_{j}) +
\sum_{i\in
A, j\in \overline{A'}} \min(x_{i}z_{j},y_{i}t_{j})  \\
&  & + \sum_{i\in \overline{A}, j\in A'} \min(x_{i}z_{j},y_{i}t_{j})
+ \sum_{i\in \overline{A}, j\in \overline{A'}}
\min(x_{i}z_{j},y_{i}t_{j}) \\
& \geq & \sum_{i\in A j\in A'} x_{i}z_{j} + \sum_{i\in A j\in \overline{A'}} x_{i}t_{j} + \sum_{i\in \overline{A} j\in A'} y_{i}z_{j} + \sum_{i\in \overline{A} j\in \overline{A'}} y_{i}t_{j} \\
& \geq & \alpha_{1}\alpha_{2} + \alpha_{1}(1- \beta_{2}) + \alpha_{2}(1 - \beta_{1}) + (1 - \beta_{1})(1 - \beta_{2}) \\
& \geq & c_1c_2
\end{eqnarray*}

By replacing the statistical closeness by the statistical
difference, we get
\[
\delta \le 1 - (1 - \delta_{1})(1 - \delta_{2})
\]
\end{proof}


\begin{proof}(Prop \ref{pola1})
Let two constants $a,b$ such that $1 > b > a > 0$.
We do this reduction in three steps:
\begin{enumerate}
\item We show that mut-$ IID^{a,b} \preccurlyeq $ mut-$IID^{\phi - \alpha,\phi + \alpha} $ with $ \alpha > 0$ and $\phi = \frac{\sqrt{5} - 1}{2}$.
\item  We show that mut-$IID^{\phi - \alpha, \phi + \alpha} \preccurlyeq $mut-$IID^{1/n^{2}, 1 - 1/n^{2}}$.
\end{enumerate}
\noindent

We show the first reduction by the following lemma:

\begin{lemma} \label{atophi}
Let $a,b$ such that $b > a$. There exists $ \alpha > 0 $ such that $ \textrm{mut -}IID^{a,b} \preccurlyeq \textrm{mut -}IID^{\phi - \alpha,\phi + \alpha} $.
\end{lemma}

\begin{proof}
Let $X,Y$ two distributions and $a,b$ with $b > a$
such that $SD(X,Y) \leq a$ or $\textrm{mut-}Disj(X,Y) \ge b$. We are going to construct a pair of distributions $(A,B)$ with the property that either $SD(A,B) \leq \phi - \alpha$ or $\textrm{mut-}Disj(A,B) \ge \phi + \alpha$. Let $\Gamma$ and $\Gamma'$ such that $\Gamma,\Gamma' \notin Im(X) \cup Im(Y)$. We define the following distribution:
\[
A_{\Gamma,u,X}(x) = \textrm{ With probability $u$ return } X(x)
\textrm{ else return } \Gamma.
\]
Similarly, we define the distributions $A_{\Gamma,u,Y}(x), A_{\Gamma',u,X}(x)$. We have
\begin{itemize}
\item $SD(X,Y) \leq a \implies SD(A_{\Gamma,u,X},A_{\Gamma,u,Y}) \leq u^{2}a + 2u(1-u) = f(u,a)$.
\item $\textrm{mut-}Disj(X,Y) \geq b \implies \textrm{mut-}Disj(A_{\Gamma,u,X},A_{\Gamma,u,Y}) \geq u^{2}b + 2u(1-u) = f(u,b)$
\item $SD(X,Y) \leq a \implies SD(A_{\Gamma,u,X},A_{\Gamma',u,Y}) \leq u^{2}a + 2u(1-u) + (1 - u)^{2} = g(u,a)$
\item $\textrm{mut-}Disj(X,Y) \geq b \implies \textrm{mut-}Disj(A_{\Gamma,u,X},A_{\Gamma',u,Y}) \geq u^{2}b + 2u(1-u) + (1 - u)^{2} = g(u,b)$
\end{itemize}

Let $\delta = (a + b)/2$. If $\delta = \phi$, then the distributions $X,Y$ already have the desired property. If $\delta > \phi$ then from the fact that the function $f$ is continuous, $f(0,\delta) = 0$ and $f(1, \delta) = \delta$, we conclude that there exists a constant $u_0 \in [0,1]$ such that $f(u_0,\delta) = \phi$.
The pair of distributions $(A_{\Gamma,u_0,X},A_{\Gamma,u_0,Y})$ has the desired property
\begin{itemize}
\item $SD(X,Y) \leq a \implies SD(A_{\Gamma,u_0,X},A_{\Gamma,u_0,Y}) \leq u_0^{2}a + 2u_0(1-u_0) = \phi - u_0^2 \frac{b-a}{2}$.
\item $\textrm{mut-}Disj(X,Y) \geq b \implies \textrm{mut-}Disj(A_{\Gamma,u_0,X},A_{\Gamma,u_0,Y}) \geq u_0^{2}b + 2u_0(1-u_0) = \phi + u_0^2 \frac{b-a}{2}$.
\end{itemize}
Similarly, for the case $\delta < \phi$ we use the distributions $(A_{\Gamma,u,X},A_{\Gamma',u,Y})$ and the function $g$.
\end{proof}

In order to show our third reduction, we need the following claim : Let $X$ and $Y$ two probability distributions. Denote $(U,V)=
XOR(X,Y)$ and let $T: P_{n} \times P_{n} \rightarrow P_{2n} \times
P_{2n}$ be the operator $T(X,Y) = (U \otimes U, V \otimes V)$.

\begin{claim} Let $(A,B) = T(X,Y)$
\[
\begin{array}{l}
SD(X,Y) \le \alpha \Rightarrow SD(A,B) \le 1 - (1- \alpha^{2})^{2} \\ \\
\textrm{mut -}Disj(X,Y) \ge \beta \Rightarrow \textrm{mut -}Disj(A,B) \ge 1 - (1- \beta^{2})^{2} \\ \\
\end{array}
\]
\end{claim}

\begin{proof}
The proof follows from our new upper bound on $SD$, the Direct
Product Lemma and the XOR Lemma.
\begin{eqnarray*}
SD(A,B) & = & SD(U \otimes U, V \otimes V) \; \leq \; 1 - (1 -
SD(U,V))^{2}
\; = \; 1 - (1 - (SD(X,Y))^2)^{2} \\
& \leq & 1 - (1- \alpha^{2})^{2}  \\
\textrm{mut-}Disj(A,B) & = & 1 - (1 -
\textrm{mut-}Disj(U,V))^{2} = \; 1 - (1 - (\textrm{mut-}Disj(X,Y))^2)^{2} \\
& \geq & 1 - (1- \beta^{2})^{2}
\end{eqnarray*}
\end{proof}

We now have:
\begin{lemma} \label{phiton}
Let $\phi = \frac{\sqrt{5} - 1}{2}.$ For any $\alpha > 0$ (constant),
$\textrm {mut-}IID^{\phi - \alpha, \phi + \alpha} \preccurlyeq IID^{1/n^{2}, 1 - 1/n^{2}}$.
\end{lemma}

\begin{proof}
Let $f(x) = 1 - (1 - x^{2})^{2}$ and $U_{i+1} = f(U_{i})$. The fixed point of $f$ is $\phi = \frac{\sqrt{5} - 1}{2}$. By a straightforward study of $f$, we can see that if $U_{0} \le \phi - \alpha$ then $U_{k} \le 1/n^{2} $ and if $U_{0} \ge \phi + \alpha$ then $U_{k} \geq  1 - 1/n^{2}$ with $k = poly(n)$.

Let $(A^{i},B^{i}) = T^{i}(X,Y)$. By the previous Claim, we know that $SD(A^{i},B^{i})$ and $\textrm{mut -}Disj(A^{i},B^{i})$ behave like $U_{i}$.
Then, for $(A,B) = T^{k}(X,Y)$ we know that the size of the final distribution is
$n \cdot 2^{u} = poly(n)$ and
\begin{eqnarray*}
SD(X,Y) \le \phi - \alpha & \Rightarrow & SD(A,B) \le 1/n^{2} \\
\textrm{mut-}Disj(X,Y) \ge \phi + \alpha & \Rightarrow & \textrm{mut-}Disj(A,B) \ge 1 - 1/ n^{2}
\end{eqnarray*}

\end{proof}

From the original polarization of \cite{BG03}, we know that $\textrm{mut -}IID^{1/n^2,1 - 1/n^2} \preccurlyeq \textrm{mut -}IID^{2^{-k}, 1 - 2^{-k}}$.
Putting these two reductions together as well as the original polarization, we have that for $1>b>a>0$:
\[
IID^{a,b'} \preccurlyeq \textrm{mut -}IID^{a,b} \preccurlyeq \textrm{mut -}IID^{\phi - \alpha, \phi + \alpha} \preccurlyeq
\textrm{mut -}IID^{1/n^{2}, 1 - 1/n^{2}} \preccurlyeq \textrm{mut -}IID^{2^{-k}, 1 - 2^{-k}}
\]
\end{proof}

\end{appendix}

\end{document}